\documentclass{article}

\usepackage{arxiv}

\usepackage[english]{babel}

\usepackage[utf8]{inputenc} 
\usepackage[T1]{fontenc}    
\usepackage[round,authoryear]{natbib}

\usepackage{lmodern}
\usepackage{amsmath,amssymb}
\usepackage{booktabs}
\usepackage{array}
\usepackage{longtable}
\usepackage{tabularx}
\usepackage{microtype}
\usepackage{xurl}
\usepackage{hyperref}
\usepackage{placeins}
\usepackage{array}

\usepackage{tikz}
\usetikzlibrary{arrows.meta, positioning, calc, shapes.geometric}

\usepackage{adjustbox}
\usepackage{pifont} 

\usepackage{authblk}
\usepackage[capitalise,nameinlink]{cleveref}

\newcolumntype{Y}{>{\raggedright\arraybackslash}X}

\begin{document}

\title{Governing Artificial Intelligence as a Macro Socio-Technical Regime: The F–I–C–P Framework for Regime-Level AI Infrastructure and Governance}

\newcommand{\shorttitle}{The F–I–C–P Framework for Regime-Level AI Infrastructure and Governance} 

\author{Francisco Herrera$^1$$^2$}
\affil{$^1$Department of Computer Science and Artificial Intelligence, Andalusian Institute of Data Science and Computational Intelligence (DaSCI), University of Granada, Spain. \\ \texttt{herrera@decsai.ugr.es} }
\affil{$^2$ ADIA Lab, Abu Dhabi,United Arab Emirates}

\maketitle

\begin{abstract}
Artificial intelligence (AI) is often analysed through technical, organisational, or regulatory perspectives that treat capability, adoption, and governance as separate issues. This article conceptualises AI instead as an emerging macro socio-technical regime generated through the interaction of finance, infrastructure, capability, and productivity. It develops the F--I--C--P framework to explain how finance enables infrastructure, infrastructure conditions capability, and capability creates the possibility--but not the guarantee--of productivity and value realisation. The article also introduces the concept of regime-level AI infrastructure to identify concrete configurations, such as AI Gigafactories, foundation-model platforms, national compute infrastructures, public AI platforms, sectoral AI ecosystems, and corporate AI infrastructures, through which the macro-regime becomes observable and evaluable. The central analytical problem is structural misalignment: the risk that financial expectations, infrastructural expansion, technical capability, organisational absorption, productivity realisation, governance capacity, and societal outcomes evolve at different speeds or under unequal conditions of power. Governance is therefore conceptualised not as a final stage after technological deployment, but as a transversal and anticipatory steering capacity. Briefly illustrated through Spain and Chile as non-frontier configurations, the article argues that AI governance should be understood as regime alignment: the task of anticipating, evaluating, and correcting structural misalignments before they become embedded in infrastructural dependencies, organisational routines, market structures, and international asymmetries.
\end{abstract}


\newpage

\section*{Highlights}

\begin{itemize}
\item Conceptualises AI as an emerging macro socio-technical regime.
\item Develops the F--I--C--P framework linking finance, infrastructure, capability, and productivity.
\item Introduces regime-level AI infrastructure as an observable and evaluable unit of AI governance.
\item Identifies structural misalignment as a central risk of regime-level AI development.
\item Reframes AI governance as transversal and anticipatory regime alignment.
\end{itemize}
\newpage

\section{Introduction}

Artificial intelligence (AI) is increasingly described as a general-purpose technology capable of transforming production, labour, scientific research, education, public administration, communication, defence, and organisational decision-making. Yet its social and economic consequences cannot be inferred from technical capability alone. Socio-technical regime theory shows that technological change becomes consequential through configurations of technologies, infrastructures, markets, institutions, user practices, expectations, and rules \citep{geels2002, geels2004, markard2012}. From this perspective, AI should not be analysed merely as a set of models, applications, or tools, but as an emerging macro socio-technical regime.

The central difficulty is that the main dimensions of this regime are often examined separately. Technical capability is analysed apart from the infrastructures that enable it; infrastructure is examined apart from the financial commitments that build and discipline it; productivity is discussed apart from the organisational and institutional conditions required to realise it; and governance is frequently treated as a regulatory layer added after technological deployment. This separation obscures the structural relations through which AI becomes socially, economically, and politically consequential. Existing research on AI productivity, labour, infrastructure, supply chains, and governance provides important insights into these dimensions, but rarely integrates them into a single regime-level architecture \citep{acemoglu2025, acemogluJohnson2023, delRioChanona2025, kwonPorter2025, muldoon2026, valdivia2025, ghosh2025}.

This article addresses that gap by developing the F--I--C--P framework
\((F \rightarrow I \rightarrow C \rightarrow P)\), which conceptualises finance,
infrastructure, capability, and productivity as the generative architecture of
the AI regime.

Finance mobilises and allocates capital; infrastructure converts capital into material and computational capacity; capability defines what AI systems can potentially do; and productivity represents the realisation of economic, organisational, or public value. This framing builds on research showing that AI's productivity effects depend not only on technical progress, but also on task composition, organisational complementarity, labour institutions, and the creation of socially useful new tasks \citep{acemoglu2025, acemogluJohnson2023, alam2024, li2024, sohn2025}.

The article further introduces the concept of regime-level AI infrastructure. The macro-regime perspective does not imply that AI exists only as an abstract global system. Rather, the regime is materialised through concrete configurations in which financial commitments, computational infrastructure, capability production, organisational dependence, productivity expectations, governance arrangements, and societal consequences become structurally interconnected. These configurations may include AI Gigafactories, foundation-model platforms, national compute infrastructures, public AI platforms, sectoral AI ecosystems, and corporate AI infrastructures. Recent work has already argued that AI governance must increasingly confront AI not only as a tool, but also as infrastructure \citep{naser2026}; infrastructure studies similarly conceptualise infrastructure as relational, embedded, and organisationally consequential rather than as a neutral technical layer \citep{starRuhleder1996}. European AI Gigafactories and generative AI platforms illustrate how such configurations may combine compute, access, capability production, dependency, governance, and value expectations \citep{councilEU2026AIGigafactories, europeanCommissionAIGigafactories2026, nzembayieUrbano2026}.

The central problem addressed in this article is therefore structural misalignment. AI-related risks do not arise only from biased models, unsafe systems, or insufficient organisational controls. They also arise when finance expands faster than realised productivity, when infrastructure becomes concentrated in ways that restrict access to capability, when technical capability grows faster than organisational absorption, when productivity gains fail to translate into welfare or legitimacy, or when governance responds too slowly to shape trajectories before they become locked in. This interpretation is consistent with recent work on structural AI risk and with path-dependence research showing how early trajectories may become difficult to reverse once they are embedded in institutional and organisational arrangements \citep{kilian2026, pierson2000}.

The article asks: How do finance, infrastructure, capability, and productivity interact to constitute an emerging macro socio-technical AI regime; under what conditions do concrete AI configurations acquire the character of regime-level AI infrastructure; and how can governance anticipate and address the structural misalignments generated by these interactions?

The answer developed in this article is that AI governance should be understood as regime alignment: the institutional task of keeping financial expectations, infrastructural capacity, technical capability, organisational absorption, productivity realisation, governance capacity, and societal outcomes within sustainable relations. Governance is therefore not conceptualised as a final stage after technological development, but as a transversal steering capacity acting across capital allocation, infrastructure access, capability development, organisational adoption, productivity distribution, and societal consequences. This view complements recent accounts of AI governance that emphasise adaptive regulation, stakeholder accountability, regulatory lag, and international governance, while shifting the focus toward the structural relations through which AI regimes are financed, built, scaled, deployed, and evaluated \citep{cordeiro2026, ghosh2025, wang2026, emeryXu2025}.

The article makes five contributions. First, it conceptualises AI as a macro socio-technical regime rather than as a discrete technology or a collection of applications. Second, it develops the F--I--C--P framework as an analytical architecture linking finance, infrastructure, capability, and productivity. Third, it introduces the concept of regime-level AI infrastructure to identify the concrete configurations through which the macro-regime becomes observable and evaluable. Fourth, it defines structural misalignment as the central analytical problem of the AI regime. Fifth, it reconceptualises governance as a transversal and anticipatory capacity for steering the relationships among finance, infrastructure, capability, productivity, regime-level infrastructures, and societal outcomes.

The argument proceeds as follows. Section~\ref{sec:foundations} presents the conceptual foundations and analytical approach. Section~\ref{sec:architecture} defines the finance--infrastructure--capability--productivity architecture. Section~\ref{sec:governance} introduces governance, conditioning mechanisms, and societal outcomes. Section~\ref{sec:misalignments} develops the concept of regime-level AI infrastructure, identifies structural misalignments, and interprets them through structural power. Section~\ref{sec:spainchile} briefly illustrates the framework through Spain and Chile as non-frontier configurations. Section~\ref{sec:implications} derives implications for alignment, anticipation, and evaluation. Section~\ref{sec:conclusion} concludes.

\section{Conceptual Foundations and Analytical Approach}\label{sec:foundations}

This article develops an integrative conceptual framework for analysing AI as a macro socio-technical regime. Its aim is not to assess particular AI systems or to measure national AI readiness, but to explain how finance, infrastructure, capability, productivity, and governance interact as constitutive dimensions of a broader regime whose effects extend across economies, organisations, institutions, and societies.

The framework synthesises four bodies of literature: socio-technical systems and transition studies, political economy and structural power, AI productivity and labour research, and responsible and anticipatory AI governance. Each of these literatures captures an important part of the AI transformation, but none alone explains the structural interdependence among capital allocation, computational infrastructure, technical capability, productivity realisation, and societal outcomes.

\subsection{AI as a Macro Socio-Technical Regime}

Socio-technical regime theory describes relatively stabilised configurations of technologies, infrastructures, institutions, markets, user practices, regulations, and cultural expectations \citep{geels2002, geels2004, markard2012}. AI extends this perspective because it is not confined to a single sector. Its development and deployment affect finance, labour, science, logistics, education, public administration, defence, communication, and organisational decision-making. It also depends on a global material base composed of semiconductors, data centres, cloud infrastructures, energy systems, data resources, and supply chains.

AI should therefore be understood as a macro socio-technical regime: a broad configuration of technical capabilities, infrastructures, markets, institutions, expectations, and power relations that increasingly shapes the conditions under which economic and social activity is organised. This regime-level perspective is necessary because AI-related transformations cannot be explained by technical capability alone. Model performance, autonomy, and scalability matter, but they are enabled by infrastructures, financed through specific investment arrangements, embedded in organisations, and mediated by governance.

The proposed framework specifies the internal architecture of this emerging regime by identifying four generative dimensions: finance, infrastructure, capability, and productivity. These dimensions form a sequence of interdependence in which finance enables infrastructure, infrastructure conditions capability, capability creates the possibility of productivity, and expectations of productivity feed back into finance. Governance is not treated as an additional stage, but as a transversal capacity acting across these relations.

\subsection{From AI-as-Tool Governance to Regime-Level AI Infrastructure}
\label{sec:tool-to-infrastructure}

Much AI governance still begins from an AI-as-tool perspective. From this view, the central governance task is to ensure that individual AI systems are safe, accurate, transparent, accountable, fair, robust, and compliant. These requirements remain indispensable. However, they are insufficient when AI systems become embedded in the infrastructures, platforms, organisational routines, financial expectations, and institutional arrangements through which economic and social activity is organised. The governance problem then shifts from the assessment of isolated systems to the steering of structural relations.

This difference can be expressed schematically as follows:

\[
G_{\mathrm{tool}} \rightarrow
\{\text{safety},\ \text{accuracy},\ \text{fairness},\ \text{accountability},\ \text{compliance}\}.
\]

This formulation captures the logic of system-level governance. It is appropriate when AI remains a discrete application or organisational tool. However, it becomes insufficient when AI configurations begin to organise access to capability, structure organisational workflows, create dependency, concentrate value capture, or condition future productivity. In such cases, the relevant object of governance is no longer only the individual AI system, but the structural configuration through which AI becomes infrastructural.

This shift is especially important because AI does not become socially consequential through technical capability alone. AI capability depends on computational infrastructure, cloud platforms, semiconductors, data, energy systems, supply chains, specialised labour, capital allocation, organisational absorption, and governance capacity. These elements are not external background conditions; they shape who can develop AI, who can access it, who depends on it, and who captures value from it. In this sense, AI increasingly operates not only as a tool used within organisations, but also as an infrastructural condition for other forms of activity.

This argument is consistent with recent work suggesting that AI governance instruments must increasingly confront AI not only as a tool, but also as infrastructure \citep{naser2026}. It also draws on infrastructure studies that conceptualise infrastructure not as a neutral technical layer, but as a relational and embedded arrangement connecting technical systems, organisational practices, access conditions, and large-scale institutional change \citep{starRuhleder1996}. From this perspective, infrastructure is not defined by technical scale alone. It becomes infrastructural when other actors organise their practices around it, when access to it conditions participation, and when it becomes difficult to substitute without significant organisational or institutional disruption.

The concept of regime-level AI infrastructure captures this transition. Many AI systems remain discrete applications, organisational tools, or sector-specific solutions. Others acquire regime-level infrastructural character when they integrate financial commitments, computational infrastructure, capability production, organisational dependence, productivity expectations, governance arrangements, and societal consequences. This transition can be represented schematically as:

\[
\text{AI configuration}
\Rightarrow
\text{Regime-Level AI Infrastructure}
\]

when finance, infrastructure, capability, organisational dependence, productivity expectations, governance arrangements, and societal consequences become structurally interconnected.

The analytical implication is that regime-level AI infrastructures must be evaluated through a broader lens than system-level AI governance. Their risks cannot be assessed only in terms of safety, accuracy, fairness, or compliance. They must also be evaluated in terms of dependency, concentration, substitutability, lock-in, governance latency, productivity realisation, value capture, resilience, and legitimacy. In this sense:

\[
\mathrm{Risk}_{\mathrm{RLAI}} \neq \mathrm{Risk}_{\mathrm{AI\ system}}.
\]

Here, \(\mathrm{RLAI}\) denotes regime-level AI infrastructure. The expression is not intended as a formal risk model, but as a conceptual distinction between two different objects of governance: the individual AI system and the infrastructural configuration through which AI becomes embedded in wider economic, organisational, and institutional relations.

Current debates on AI productivity, labour, infrastructure, governance, and political economy reinforce this point. Studies of AI and productivity question whether rapid technical progress will generate proportional aggregate gains, particularly when automation affects only some tasks, when cost savings remain limited, or when new-task creation is insufficient \citep{acemoglu2025}. Research on work and automation emphasises that AI's effects depend on whether systems complement or substitute human labour, and on whether organisations redesign tasks, workflows, and responsibilities in ways that support augmentation rather than displacement \citep{acemogluJohnson2023, alam2024, delRioChanona2025}. At the same time, studies of AI infrastructure, supply chains, and platform power show that compute, cloud platforms, semiconductors, data, energy, and global production networks increasingly shape who can develop, deploy, and profit from advanced AI systems \citep{kwonPorter2025, muldoon2026, valdivia2025, xiao2025}.

This transition from AI-as-tool governance to regime-level AI infrastructure provides the conceptual bridge for the framework developed in the following sections. It explains why AI governance must address not only individual applications, but also the structural relationships through which AI is financed, built, scaled, adopted, embedded, and translated into productivity and societal outcomes. The next section specifies these relationships through the F--I--C--P architecture.

\subsection{Structural Power and Integrative Conceptual Analysis}

A regime-level analysis of AI also requires attention to power. The relevant question is not only what AI systems can do, but who controls the conditions under which they are financed, built, deployed, accessed, and governed. Susan Strange's theory of structural power is useful because it shifts attention from direct influence over particular decisions to the capacity to shape the frameworks within which other actors operate \citep{strange1988, strange1996}.

Applied to AI, structural power resides not only in ownership of models or data, but also in control over finance, infrastructure, supply chains, standards, cloud platforms, regulatory architectures, and productivity gains. Actors that control these conditions can shape the direction of innovation, the distribution of value, the terms of access, and the dependencies faced by others. This is especially important for non-frontier economies, which may adopt AI systems widely without controlling the financial, infrastructural, or technical foundations of the regime.

Methodologically, the article is an integrative conceptual synthesis. It does not test a hypothesis using a dataset or offer a formal dynamic model. Instead, it constructs an analytical framework by bringing together concepts, mechanisms, and empirical tendencies identified across several research streams. The resulting propositions are conceptual rather than empirically validated claims; they are intended to clarify relationships that future research can operationalise and test.

The article also uses Spain and Chile as illustrative international configurations. These examples demonstrate the usefulness of the framework for understanding non-frontier positions within the AI regime, but they are not presented as exhaustive comparative case studies. The conceptual contribution is therefore twofold: first, the article identifies the F--I--C--P architecture as the core generative structure through which AI becomes socially and economically consequential; second, it reconceptualises governance as a transversal steering capacity directed toward structural misalignments. This framing is necessary because the societal effects of AI cannot be explained by capability, adoption, or regulation in isolation. They emerge from the interaction between capital, infrastructure, technical capability, productivity realisation, institutional capacity, and structural power.

This framing contributes to technology-in-society research by explaining how technological change becomes socially consequential through the interaction of economic systems, material infrastructures, organisational practices, governance capacity, and distributive outcomes. AI is therefore analysed not as an autonomous driver of social transformation, but as a regime whose consequences depend on the institutional and political-economic structures through which it is financed, built, adopted, and governed.

\section{Finance, Infrastructure, Capability, and Productivity: The Generative Architecture of the AI Regime}\label{sec:architecture}

The previous section established why AI governance must move from the assessment of isolated systems to the analysis of regime-level infrastructures and structural relations. This section specifies the generative architecture of those relations. Its purpose is not to propose a deterministic sequence or a formal dynamic model, but to identify the structural order through which AI becomes economically, organisationally, and socially consequential: capital is mobilised, infrastructure is built, capability is produced, and value may---or may not---be realised.

The proposed framework conceptualises AI as a macro socio-technical regime generated through the interaction of four structural dimensions: finance, infrastructure, capability, and productivity.

At its simplest, this architecture can be represented as:

\begin{equation}
F \rightarrow I \rightarrow C \rightarrow P.
\end{equation}

Finance enables the expansion of infrastructure; infrastructure conditions the production and deployment of capability; capability creates the technical possibility of productivity; and productivity, in turn, sustains or challenges the expectations upon which further investment depends. The notation is deliberately simple. It is not an econometric model, but an analytical grammar for relating dimensions that are often studied separately.

This distinction is central because technical capability does not automatically produce economic or social value. AI systems may expand the range of automatable tasks, but their translation into productivity depends on infrastructure, organisational adaptation, institutional direction, labour arrangements, and the creation of complementary forms of work. Similarly, infrastructure expansion may accelerate technical capability while generating financial exposure if expected returns fail to materialise within anticipated timeframes. The core analytical problem is therefore not the growth of any single dimension, but the alignment---or misalignment---among them.

This section defines the four generative dimensions of the regime. It first examines finance as the mechanism through which capital is mobilised and risk is allocated. It then analyses infrastructure as the material and organisational substrate of AI capability. It next defines capability as technical potential rather than realised value. Finally, it conceptualises productivity as value realisation and introduces the feedback loop through which expectations of future productivity reproduce or destabilise the regime.

\subsection{Finance}

Finance refers to the mechanisms through which capital is mobilised, allocated, priced, and exposed to risk in support of AI development. These mechanisms include venture capital, strategic equity, corporate investment, public funding, debt, private credit, special-purpose vehicles, securitisation, and other financial arrangements used to support the expansion of computational infrastructure.

In the AI regime, finance does not merely provide external support for technological development. It shapes the scale, speed, direction, ownership, and risk profile of technological expansion. Decisions about capital allocation influence which infrastructures are built, where they are located, who controls them, what returns are expected, and how risks are distributed across firms, investors, states, and wider economic systems.

This is particularly important because advanced AI development and deployment require large-scale and long-term investment in semiconductors, data centres, cloud systems, energy networks, connectivity, data resources, and specialised technical capacity. These investments are often justified by expectations of future productivity, market dominance, revenue growth, or strategic advantage. Finance therefore becomes structurally linked to expectations about the future performance of AI systems and the future value they may generate.

The finance--infrastructure relationship can be expressed as:
\begin{equation}
F \rightarrow I.
\end{equation}
Finance enables infrastructure, but it also disciplines it. It determines not only how much infrastructure is built, but also what forms of infrastructure are prioritised, which actors gain control over the material foundations of AI capability, and which risks are shifted across firms, investors, states, workers, and users.

This relationship also introduces a central source of fragility into the AI regime. If expected productivity exceeds realised productivity over sustained periods, financial commitments may support infrastructure expansion that is economically, socially, or institutionally difficult to justify. In such conditions, the regime may become exposed to overinvestment, debt pressure, valuation instability, and pressure to accelerate automation in order to validate prior investment expectations.

Finance should therefore be understood as a constitutive dimension of the AI regime, not as a background condition. It links expectations, capital allocation, infrastructure expansion, ownership structures, and risk distribution. The question is not simply whether AI attracts investment, but how that investment shapes the material conditions of capability, the concentration of control, and the future trajectory of productivity and social value.

\subsection{Infrastructure}

Infrastructure denotes the material and organisational substrate through which financial investment is converted into operational AI capacity. It includes specialised semiconductors, high-performance computing systems, data centres, cloud platforms, energy networks, telecommunications infrastructure, data resources, supply chains, and the specialised technical labour required to maintain and scale these systems.

This understanding is consistent with infrastructure studies, which conceptualise infrastructure not as a neutral technical layer, but as a relational and embedded arrangement connecting technical systems, organisational practices, access conditions, and large-scale institutional change \citep{starRuhleder1996}.

Within the AI regime, infrastructure performs two simultaneous functions. First, it acts as a productive input: without sufficient computational capacity, energy, connectivity, and data availability, technical capability cannot be developed, trained, deployed, or scaled. Second, infrastructure operates as a source of structural power. Control over computational resources conditions who can develop advanced models, who can deploy them at scale, who can access downstream markets, and who remains dependent on external providers.

This dual function makes infrastructure more than a neutral technical layer. Control over computational resources has become a central locus of economic concentration, enabling rent extraction and shaping competition at a systemic scale.  \citet{kwonPorter2025} show how exclusive access to data and machine-learning resources can affect innovation and competition. Infrastructure therefore determines not only the technical frontier of AI development, but also the distribution of access, dependency, and market power.

The infrastructural basis of AI is also embedded in global supply chains. Advanced AI systems depend on semiconductors, data-centre construction, energy systems, rare minerals, data labour, model deployment infrastructures, and waste streams distributed across multiple jurisdictions. These chains are politically contested and socially uneven, generating dependencies that extend beyond the immediate technical system \citep{muldoon2026}. From this perspective, AI infrastructure participates in broader dynamics of supply-chain capitalism, where environmental costs, labour extraction, and infrastructural dependencies are distributed asymmetrically across global production networks \citep{valdivia2025}. The environmental dimension is particularly relevant in the expansion of data centres, where energy demand, server location, and decarbonisation pathways shape the social and ecological costs of scaling AI systems \citep{xiao2025}.

The infrastructure--capability relationship can therefore be expressed as:
\begin{equation}
I \rightarrow C.
\end{equation}
Infrastructure conditions capability because the development and deployment of advanced AI systems require access to computational scale, energy, data, and technical platforms. At the same time, infrastructural concentration limits who can participate in frontier development and who is positioned primarily as an adopter, client, or rule-taker. In this sense, infrastructure acts as a gatekeeping layer within the AI regime: it determines what can be built, who can build it, and under what conditions it can be deployed.

\subsection{Capability}

Capability refers to the technical capacities of AI systems, including their performance, generality, autonomy, scalability, capacity for tool use, and ability to operate across multiple domains. It should not be understood merely as the automation of isolated tasks, but as the expansion of systems capable of performing cognitive, analytical, generative, and increasingly agentic functions at scale.

Capability is central to the AI regime because it defines the range of possible applications and transformations. More capable systems can automate a wider set of tasks, support more complex decision processes, interact with tools and digital environments, and potentially operate with increasing autonomy. Following \citet{amodei2026}, such developments may introduce actor-like properties that shift AI-related risk from the purely technical domain toward systemic and societal levels.

However, capability should not be confused with productivity, welfare, or social value. A system may perform well on technical benchmarks, automate specific tasks, or exceed human performance in selected domains without generating proportional economic or social benefits. Capability represents technological potential; its consequences depend on how it is embedded in organisations, infrastructures, markets, and governance arrangements.

This distinction is essential for avoiding technological determinism. The same technical capability may be used to augment human labour, support professional judgement, create new tasks, or improve public services. It may also be used to substitute workers, deskill occupations, concentrate returns, or externalise risks. The direction of technological change is therefore not determined by capability alone, but by the institutional, organisational, and economic structures that channel its deployment.

The capability--productivity relationship can be represented as:
\begin{equation}
C \rightarrow P.
\end{equation}
Yet this relationship is conditional rather than automatic. Capability creates the possibility of productivity, but it does not guarantee value realisation. The translation from $C$ to $P$ depends on absorptive capacity, organisational complementarity, task structure, labour institutions, and governance. Without these conditions, increasing capability may expand automation without producing sustainable productivity gains or broadly distributed benefits.

\subsection{Productivity and Value Realisation}

Productivity refers here to the realisation of economic, organisational, and public value from AI deployment. It should not be reduced to the mere automation of tasks or the technical efficiency of individual systems. Rather, productivity concerns whether AI capabilities generate measurable improvements in output, quality, innovation, cost reduction, organisational performance, service delivery, or the creation of socially useful new tasks.

For this reason, the framework defines $P$ as productivity and value realisation, not as welfare. This distinction is conceptually important. Productivity gains may occur without producing equitable distribution, improved job quality, institutional legitimacy, resilience, or collective well-being. Conversely, some socially valuable uses of AI may generate public value that is not immediately captured by conventional productivity metrics. The central analytical issue is therefore how technical capability becomes productive value, and under what conditions that value contributes to broader social outcomes.

This distinction is consistent with Acemoglu's account of the macroeconomics of AI. The aggregate impact of AI depends not simply on technical progress, but on the share of tasks affected, the real cost savings generated, the distinction between easy-to-learn and hard-to-learn tasks, and the creation of new tasks \citep{acemoglu2025}. In the absence of complementary organisational changes and socially useful task creation, technical capability may generate limited aggregate productivity gains even under conditions of rapid automation.

The productivity paradox reinforces this point. Advances in algorithmic capability and large-scale investment in AI have not yet translated clearly into sustained increases in aggregate productivity \citep{sohn2025}. This does not necessarily imply that AI is economically insignificant. It may indicate that capability is expanding faster than organisations, labour markets, institutions, and measurement systems can adapt. Sohn's notion of ``effortlessness'' is useful in this context because it highlights how AI may assist, reduce, or replace human effort in ways that are not fully captured by conventional productivity indicators \citep{sohn2025}.

The relationship between capability and productivity also depends on whether AI systems complement or substitute human labour. \citet{delRioChanona2025} show that the employment and productivity effects of AI vary according to the types of tasks affected and whether systems augment or replace human workers. Augmentation-oriented configurations may preserve professional judgement while supporting productivity gains, as illustrated by work on NextGen-AI in engineering and manufacturing \citep{alam2024}. Organisational learning and effective use also depend on task complexity and user engagement rather than technical capability alone \citep{li2024}.

This perspective connects directly with the broader argument advanced by \citet{acemogluJohnson2023}: technological innovation does not automatically generate shared prosperity. Its effects depend on whether institutions and organisational arrangements favour complementarity, new-task creation, and broadly distributed value, or whether they favour substitution, rent concentration, and labour displacement.

Productivity should therefore be understood as an achievement of socio-technical integration. The relevant question is not only whether AI systems become more capable, but whether those capabilities are embedded in productive arrangements that create durable value. In this sense:
\begin{equation}
P \neq C
\end{equation}
and:
\begin{equation}
P = \Phi(F, I, C, G \mid A, O, H, T).
\end{equation}
This expression is not intended as an econometric model, but as a conceptual representation of the fact that productivity depends on capability only under specific infrastructural, financial, organisational, institutional, and temporal conditions. Capability is a technological potential; productivity is an organisational and institutional achievement.

\subsection{Expectations, Feedback, and Regime Reproduction}

The AI regime is not a simple linear sequence from finance to productivity. It is a recursive structure in which expectations about future productivity feed back into present financial commitments. Financial actors invest in infrastructure because they anticipate future revenue, productivity gains, market dominance, or strategic advantage. These expectations then shape the scale and speed of infrastructural expansion, which in turn conditions the development of technical capability.

The basic architecture can be expressed as:
\begin{equation}
F \rightarrow I \rightarrow C \rightarrow P.
\end{equation}
However, the regime reproduces itself through a feedback loop:
\begin{equation}
P^{e}_{t+1} \rightarrow F_t,
\end{equation}
where $P^{e}$ denotes expected productivity or expected future value. This distinction between expected productivity and realised productivity is crucial. Expected productivity supports investment narratives, market valuations, infrastructure expansion, and strategic commitments. Realised productivity reflects the actual capacity of AI systems to generate economic, organisational, or public value once deployed.

The relationship can therefore be expressed as a tension between:
\begin{equation}
P^{e} = \text{expected productivity}
\end{equation}
and:
\begin{equation}
P^{r} = \text{realised productivity}.
\end{equation}
When $P^{e}$ substantially and persistently exceeds $P^{r}$, the regime may become vulnerable to structural misalignment. Infrastructure may expand faster than productive use cases mature. Financial obligations may depend on returns that are delayed, uncertain, or unevenly distributed. Firms may intensify automation to justify investment. States may support infrastructure expansion without sufficient capacity to ensure broad social value. In such conditions, the finance--infrastructure--capability cycle may generate fragility rather than sustainable transformation.

This feedback dynamic is especially important because AI infrastructure is capital intensive and often concentrated in a small number of firms and jurisdictions. If expected returns fail to materialise within anticipated timeframes, financial exposure can accumulate around infrastructure, valuations, debt obligations, and market expectations. Conversely, if realised productivity is broadly distributed and institutionally supported, the cycle may reinforce sustainable investment, organisational learning, and social value creation.

This feedback loop is the point at which technological expectation becomes a social and economic force. Expectations about future AI productivity do not merely describe possible outcomes; they organise present investment, infrastructure expansion, competitive behaviour, and institutional pressure. For this reason, the governance of AI must also be understood as the governance of expectations, not only the governance of deployed systems.

The F--I--C--P architecture therefore captures both the generative power and the fragility of the AI regime. Finance enables infrastructure; infrastructure enables capability; capability enables the possibility of productivity; and expected productivity sustains finance. But the stability of the regime depends on whether these dimensions remain aligned. The following section extends this architecture by introducing governance as a transversal steering capacity, the conditioning mechanisms that mediate the translation from capability to productivity, and the societal outcomes through which the regime must ultimately be assessed.

\section{Governance, Conditioning Mechanisms, and Societal Outcomes}\label{sec:governance}

The F--I--C--P architecture explains how the AI regime is generated, but not how its social trajectory is steered. Finance enables infrastructure, infrastructure conditions capability, and capability creates the possibility of productivity; however, this sequence does not determine whether AI produces broad welfare, fair distribution, resilience, or legitimacy. Similar levels of investment, infrastructure, or capability may generate very different outcomes depending on institutional capacity, organisational absorption, structural concentration, and the timing of governance responses.

For this reason, the framework requires a second analytical layer: governance, conditioning mechanisms, and societal outcomes. Governance is treated not as a final stage after productivity, but as a transversal steering capacity acting across the regime. Conditioning mechanisms explain why similar technologies produce different productivity effects across countries, sectors, and organisations. Societal outcomes distinguish productivity from the broader consequences through which the regime must ultimately be assessed.

The overall  F--I--C--P architecture of the AI macro socio-technical regime is summarised in Figure~\ref{fig:ficp-regime}.

\subsection{Governance as Transversal Steering Capacity}

Governance should not be conceptualised as the final element of a linear sequence:
\begin{equation}
F \rightarrow I \rightarrow C \rightarrow P \rightarrow G.
\end{equation}
Such a formulation would imply that governance intervenes only after finance, infrastructure, capability, and productivity have already developed. This would reproduce the limits of reactive regulation, in which institutions respond to technological trajectories once they are already economically embedded or politically difficult to reverse.

Instead, governance operates transversally across the AI regime:
\begin{equation}
G \leftrightarrow (F, I, C, P).
\end{equation}
Governance shapes the conditions under which capital is allocated, infrastructure is built, capability is developed, and productivity is distributed. It also receives feedback from the regime in the form of financial fragility, infrastructural dependency, labour disruption, social distrust, or geopolitical asymmetry. Governance is therefore both a constitutive condition of the AI regime and an adaptive response to its consequences.

In this sense, governance is not merely regulation. It includes law, technical standards, institutional coordination, auditing systems, accountability mechanisms, competition policy, public investment, procurement, safety requirements, data governance, organisational procedures, and international cooperation. Responsible AI governance remains necessary for transparency, accountability, safety, fairness, robustness, and human oversight, but regime-level governance must also address the structural interactions through which finance, infrastructure, capability, and productivity generate systemic effects.

This regime-level view complements recent accounts of AI governance that emphasise adaptive regulatory approaches, transparency, bias, and the economic influence of increasingly complex AI systems, while shifting the focus toward the structural relations through which finance, infrastructure, capability, and productivity interact \citep{ghosh2025}.

Effective governance depends on four institutional capacities. Observability enables institutions to detect changes in investment flows, infrastructure concentration, technical capabilities, deployment patterns, labour-market effects, productivity outcomes, and systemic dependencies. Interpretability enables them to understand how developments in one dimension affect the others. Controllability provides instruments capable of altering trajectories, including regulation, financial supervision, competition policy, public infrastructure, procurement, compute access rules, auditing, and technical standards. Adaptability enables institutions to revise these instruments as technologies, markets, and risks evolve. Governance effectiveness therefore depends not merely on the existence of rules, but on the capacity to observe, interpret, influence, and adapt to the evolving regime.

\subsection{Conditioning Mechanisms}

The movement from capability to productivity is not automatic. Between the generative architecture of the regime and its realised outcomes lie conditioning mechanisms that explain why similar technologies may produce different effects across countries, sectors, organisations, and institutional contexts.

The first mechanism is absorptive capacity. This refers to the ability of an organisation, sector, or country to recognise, assimilate, adapt, and productively use AI capabilities. It includes human capital, technical skills, managerial competence, research ecosystems, institutional learning, administrative capacity, and the diffusion of innovation across firms and public organisations.

The second mechanism is organisational complementarity. This concerns the redesign of tasks, workflows, roles, incentives, training, and accountability structures required for AI to complement rather than simply substitute human labour. The distinction between augmentation and substitution is not inherent in the technology itself; it depends on organisational choices and institutional incentives.

The third mechanism is structural concentration. Concentration refers to the distribution of control over capital, compute, cloud platforms, data, models, intellectual property, standards, and market access. It affects the finance--infrastructure relation, the infrastructure--capability relation, and the distribution of productivity gains.

The fourth mechanism is temporal adaptation and institutional latency. Financial expectations, infrastructure expansion, technical capability, organisational absorption, productivity realisation, and governance capacity evolve at different speeds. These differential speeds generate misalignments when institutions cannot adapt quickly enough to shape technological and financial trajectories before they become locked in.

These four mechanisms can be expressed conceptually as:
\begin{equation}
P = \Phi(F, I, C, G \mid A, O, H, T),
\end{equation}
where $A$ denotes absorptive capacity, $O$ organisational complementarity, $H$ structural concentration, and $T$ temporal adaptation. This expression is not intended as an estimated function. It clarifies that productivity and value realisation depend not only on capability, but on the institutional and organisational conditions under which capability is financed, deployed, absorbed, and governed.

\subsection{Societal Outcomes}

The final analytical distinction concerns the difference between productivity and societal outcomes. Productivity and value realisation indicate whether AI capabilities generate economic, organisational, or public value. However, they do not determine how that value is distributed, whether it improves welfare, whether the regime is resilient, or whether institutions retain legitimacy and social trust.

The framework therefore distinguishes four societal outcomes. Welfare refers to the broader social value generated by AI, including income, job quality, access to services, public value, opportunities, and collective well-being. Distribution and dependency concern who captures the benefits of AI and who remains dependent on external providers for compute, cloud services, models, capital, standards, or technical knowledge. Resilience refers to the capacity of countries, organisations, and institutions to withstand financial, infrastructural, geopolitical, environmental, operational, or social shocks affecting the AI regime. Legitimacy and trust concern whether AI deployment is perceived as fair, accountable, secure, justifiable, and socially beneficial.

The relationship between regime architecture, governance, conditioning mechanisms, and societal outcomes can therefore be represented as:
\begin{equation}
(W, D, R, L) = \Psi(F, I, C, P, G \mid A, O, H, T),
\end{equation}
where $W$ denotes welfare, $D$ distribution and dependency, $R$ resilience, and $L$ legitimacy and trust. This expression captures the central normative claim of the framework: the social consequences of AI depend not only on the expansion of capability or productivity, but on the governance of the structural conditions through which finance, infrastructure, capability, and productivity interact.

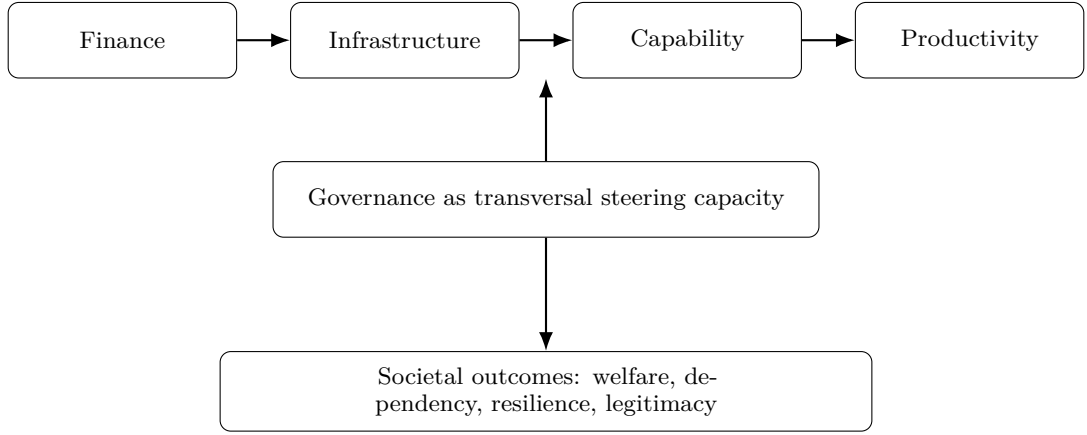
\begin{figure}[htbp]
\centering
\begin{tikzpicture}[
    node distance=0.78cm and 0.7cm,
    every node/.style={font=\small},
    box/.style={
        draw,
        rounded corners,
        align=center,
        minimum height=1cm,
        text width=2.6cm,
        inner sep=6pt
    },
    widebox/.style={
        draw,
        rounded corners,
        align=center,
        minimum height=1cm,
        text width=6.8cm,
        inner sep=6pt
    },
    outcomebox/.style={
        draw,
        rounded corners,
        align=center,
        minimum height=1cm,
        text width=8.2cm,
        inner sep=6pt
    },
    >={Latex[length=2.5mm]}
]

\node[box] (finance) {Finance};
\node[box, right=of finance] (infra) {Infrastructure};
\node[box, right=of infra] (capability) {Capability};
\node[box, right=of capability] (productivity) {Productivity};

\node[widebox, below=1.6cm of $(infra)!0.5!(capability)$] (governance)
{Governance as transversal steering capacity};

\node[outcomebox, below=1.5cm of governance] (outcomes)
{Societal outcomes: welfare, dependency, resilience, legitimacy};

\draw[->, thick] (finance) -- (infra);
\draw[->, thick] (infra) -- (capability);
\draw[->, thick] (capability) -- (productivity);

\draw[->, thick] (governance.north) -- ($(infra.south)!0.5!(capability.south)$);
\draw[->, thick] (governance.south) -- (outcomes.north);

\end{tikzpicture}
\caption{The F--I--C--P architecture of the AI macro socio-technical regime.}
\label{fig:ficp-regime}
\end{figure}

Figure~\ref{fig:ficp-regime} summarises the proposed  F--I--C--P architecture of the AI macro socio-technical regime. It shows the generative architecture of the AI regime, the feedback loop through which expected productivity sustains present financial commitments, the transversal role of governance, and the conditioning mechanisms through which capability is translated into productivity and societal outcomes.

This distinction also clarifies why the central analytical problem of the AI regime is not simply technological risk, but structural misalignment. When finance, infrastructure, capability, productivity, governance, and societal outcomes evolve at different speeds or under unequal conditions of power, the regime may generate vulnerabilities that cannot be understood by examining any single dimension in isolation. The following section develops this problem through a set of analytical propositions.

\section{Regime-Level AI Infrastructure, Structural Misalignments, and Power}
\label{sec:misalignments}

The preceding sections have defined the generative architecture of the AI regime and the second-order mechanisms through which it is governed, conditioned, and translated into societal outcomes. This section develops the central analytical implication of the framework in three steps. First, it clarifies how the macro socio-technical AI regime is materialised through concrete regime-level AI infrastructures. Second, it identifies the structural misalignments that may arise among finance, infrastructure, capability, productivity, governance, and societal outcomes. Third, it interprets these misalignments through the lens of structural power.

This sequence is important because the AI regime should not be understood only as an abstract global system. It is instantiated through concrete configurations such as AI Gigafactories, foundation-model platforms, national compute infrastructures, public AI platforms, sectoral AI ecosystems, and corporate AI infrastructures. These configurations become analytically significant when they integrate capital commitments, computational infrastructure, capability production, organisational dependence, productivity expectations, governance arrangements, and societal consequences. Once such configurations acquire regime-level infrastructural character, their risks cannot be assessed only through safety, accuracy, or compliance. They must also be evaluated in terms of dependency, concentration, substitutability, productivity realisation, governance latency, and structural misalignment.

\subsection{Regime-Level AI Infrastructure: Concrete Configurations of the AI Regime}

The macro-regime perspective does not imply that every AI system constitutes a regime-level configuration. Many AI systems remain discrete applications, organisational tools, or sector-specific solutions. A system, platform, facility, or corporate ecosystem becomes analytically relevant as regime-level AI infrastructure when it integrates several structural dimensions of the AI regime: large-scale financial commitments, material and computational infrastructure, technical capability production, organisational dependence, productivity expectations, governance requirements, and societal consequences.

European AI Gigafactories provide a paradigmatic illustration of this distinction. They are not merely technical facilities for running AI models. They are large-scale infrastructural configurations formally defined as facilities capable of handling the complete lifecycle from development to large-scale inference of very large AI models and applications, supported by AI-optimised computing capacity, data-centre infrastructure, high-capacity storage and networking, secure cloud access environments, and specialised AI-oriented support services \citep{councilEU2026AIGigafactories}. The Commission also frames them as a means to provide EU innovators, researchers, industry, SMEs, and the public sector with access to large-scale computing power and high-quality data, while supporting productivity, innovation, technological autonomy, and resilience \citep{europeanCommissionAIGigafactories2026}.

The point is not that an AI Gigafactory exhausts the AI regime. Rather, it functions as an observable regime-level unit: a configuration in which finance, infrastructure, capability, productivity expectations, governance arrangements, and societal outcomes become structurally interconnected. This makes the concept useful not only for analysing global AI dynamics, but also for evaluating concrete AI infrastructures before their dependencies, lock-ins, asymmetries, and governance constraints become difficult to reverse.

The same logic applies to foundation-model platforms and generative AI ecosystems when they provide non-substitutable, platform-governed capabilities that enable organisational or entrepreneurial activity while simultaneously generating dependence and exposure to governance shocks \citep{nzembayieUrbano2026}.

The concept is also relevant for national and sectoral AI strategies. A national AI strategy should not be assessed only by the number of AI applications it promotes, the existence of policy documents, or general levels of adoption, but by the extent to which it enables, governs, or depends on regime-level AI infrastructures across critical domains. These may include scientific AI infrastructures, industrial AI infrastructures, public-sector AI infrastructures, critical-services AI infrastructures, and defence and security AI infrastructures. In each domain, the relevant question is whether AI initiatives remain discrete tools or become infrastructural configurations that organise access to capability, create dependency, structure governance, and condition future value creation.

Table~\ref{tab:regime-level-ai-infrastructure} translates this distinction into diagnostic conditions. These conditions are not intended as a rigid checklist or a fixed threshold. Rather, they identify cumulative indicators through which an AI system, facility, platform, or organisational ecosystem may acquire regime-level infrastructural character. The more strongly these conditions are present, the more appropriate it becomes to evaluate the configuration not merely as an AI application, but as an infrastructural unit of the broader AI regime.

\begin{table}[!htbp]
\centering
\caption{Conditions under which an AI system or project acquires regime-level infrastructural character.}
\label{tab:regime-level-ai-infrastructure}
\small
\renewcommand{\arraystretch}{1.15}
\begin{tabularx}{\textwidth}{p{0.28\textwidth}X}
\toprule
\textbf{Condition} & \textbf{Regime-level question} \\
\midrule

Financial intensity &
Does the system or project depend on large-scale investment, expected returns, debt, public funding, private capital, or strategic financial commitments? \\

Infrastructural embeddedness &
Does it require compute, cloud platforms, chips, data centres, energy systems, networks, data resources, and supply chains? \\

Capability production &
Does it merely use AI, or does it produce, train, scale, distribute, or mediate access to AI capabilities? \\

Organisational dependence &
Do firms, sectors, public administrations, researchers, developers, or critical workflows begin to depend on it? \\

Productivity expectation &
Is the investment justified by expected productivity, innovation, efficiency, market advantage, public value, or technological sovereignty? \\

Governance relevance &
Does it require rules for access, standards, auditing, safety, accountability, competition, procurement, interoperability, or public oversight? \\

Societal consequence &
Does it affect welfare, distribution, dependency, resilience, labour conditions, legitimacy, or strategic autonomy? \\

\bottomrule
\end{tabularx}
\end{table}

The diagnostic value of these conditions lies in their relational character. No single condition is sufficient on its own. A large investment project may remain a technical facility if it does not create organisational dependence, governance relevance, or societal consequences. Conversely, a platform or product may acquire infrastructural significance when it mediates access to capability, becomes difficult to substitute, structures organisational workflows, or creates durable dependency.

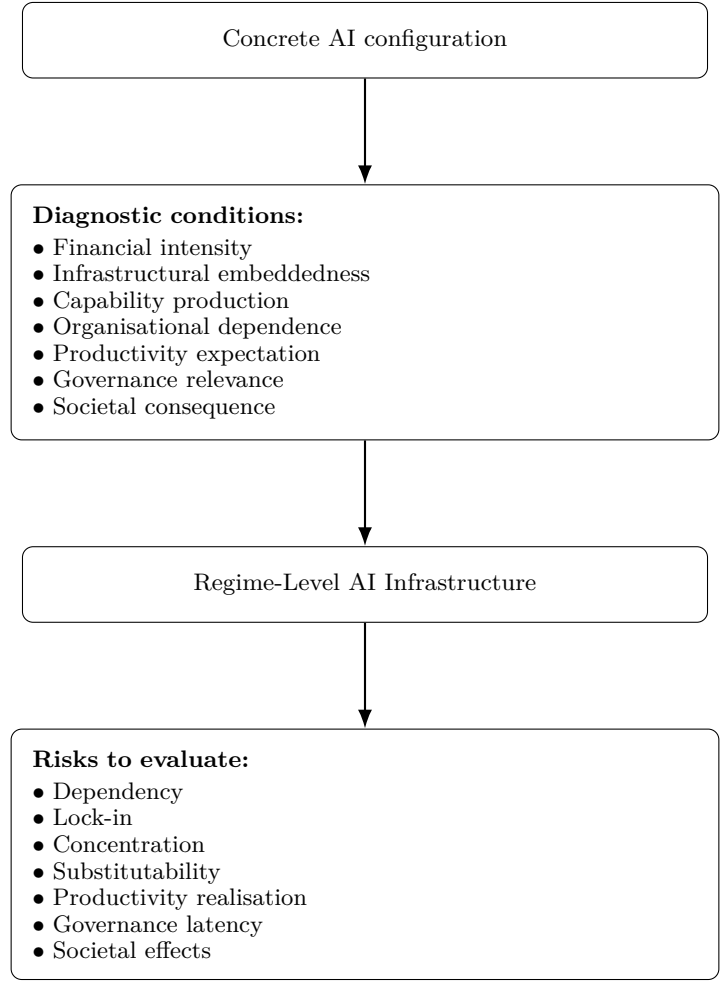
\begin{figure}[htbp]
\centering
\begin{tikzpicture}[
    node distance=1.4cm,
    every node/.style={font=\small},
    mainbox/.style={
        draw,
        rounded corners,
        align=center,
        text width=8.5cm,
        inner sep=8pt,
        minimum height=1cm
    },
    listbox/.style={
        draw,
        rounded corners,
        align=left,
        text width=8.8cm,
        inner sep=8pt
    },
    >={Latex[length=2.5mm]}
]

\node[mainbox] (config) {Concrete AI configuration};

\node[listbox, below=of config] (conditions) {
\textbf{Diagnostic conditions:}\\[2pt]
$\bullet$ Financial intensity\\
$\bullet$ Infrastructural embeddedness\\
$\bullet$ Capability production\\
$\bullet$ Organisational dependence\\
$\bullet$ Productivity expectation\\
$\bullet$ Governance relevance\\
$\bullet$ Societal consequence
};

\node[mainbox, below=of conditions] (regime) {Regime-Level AI Infrastructure};

\node[listbox, below=of regime] (risks) {
\textbf{Risks to evaluate:}\\[2pt]
$\bullet$ Dependency\\
$\bullet$ Lock-in\\
$\bullet$ Concentration\\
$\bullet$ Substitutability\\
$\bullet$ Productivity realisation\\
$\bullet$ Governance latency\\
$\bullet$ Societal effects
};

\draw[->, thick] (config) -- (conditions);
\draw[->, thick] (conditions) -- (regime);
\draw[->, thick] (regime) -- (risks);

\end{tikzpicture}
\caption{From AI systems to regime-level AI infrastructure: diagnostic conditions and evaluative dimensions.}
\label{fig:regime-level-ai-infrastructure}
\end{figure}

Figure~\ref{fig:regime-level-ai-infrastructure} summarises the diagnostic logic of the framework by showing how concrete AI configurations may acquire regime-level infrastructural character and become evaluable in terms of dependency, concentration, lock-in, substitutability, productivity realisation, governance latency, and broader societal effects.

This diagnostic move prepares the analysis of structural misalignment. Once an AI configuration acquires regime-level infrastructural character, misalignment is no longer a peripheral risk. It becomes a central governance problem because tensions among finance, infrastructure, capability, productivity, governance, and societal outcomes may become embedded in durable dependencies, investment commitments, organisational routines, market structures, and institutional constraints.

Regime-level AI infrastructure is therefore defined  by the structural integration of finance, infrastructure, capability, productivity expectations, governance, and societal outcomes. This view draws on infrastructure studies that understand infrastructure as relational, embedded, and organisationally consequential \citep{starRuhleder1996}, while also recognising that contemporary AI platforms can create enablement--dependence dynamics through non-substitutable, platform-governed capabilities \citep{nzembayieUrbano2026}. Because such configurations may generate increasing returns, institutional sequencing, and lock-in, they should be evaluated before dependencies and governance constraints become difficult to reverse \citep{pierson2000}.

\FloatBarrier

\subsection{Structural Misalignments and Analytical Propositions}

Once concrete AI configurations acquire regime-level infrastructural character, the central analytical problem becomes structural misalignment. A structural misalignment arises when one dimension of the AI regime creates demands, expectations, dependencies, or risks that the others cannot absorb, support, or govern. The issue is not that finance, infrastructure, capability, productivity, and governance must evolve at the same speed, but that their interaction may become institutionally, economically, or socially unsustainable.

This interpretation is consistent with recent work on structural AI risk, which shifts attention from discrete accidents or misuse toward system-level dynamics, feedback loops, power asymmetries, and governance leverage points \citep{kilian2026}.

The propositions should not be read as isolated hypotheses, but as a regime-level diagnostic. Each proposition identifies a point at which technological change becomes socially consequential because one dimension of the regime advances faster than, or becomes structurally detached from, the others. Together, they specify the analytical mechanism through which AI development may generate systemic vulnerability even when individual systems appear technically effective.

The first misalignment concerns the relationship between finance and infrastructure. The expansion of AI infrastructure depends increasingly on large-scale capital commitments to semiconductors, data centres, cloud systems, energy networks, and specialised technical capacity. These investments are often justified by expectations of future productivity, revenue growth, or market dominance. However, if realised productivity does not materialise within anticipated timeframes, infrastructure expansion may become financially fragile. The result is not simply overinvestment, but the conversion of technological uncertainty into financial exposure. This relationship supports the first proposition:

\paragraph{Proposition 1---Finance--Infrastructure Misalignment} The expansion of AI infrastructure depends on financial structures whose sustainability is conditional on future productivity and revenue expectations; when expected returns exceed realised value, infrastructural growth may generate systemic financial vulnerability.

The second misalignment concerns infrastructure and capability. Advanced AI capability depends on access to computational infrastructure, energy, cloud systems, data, and supply chains. Where these resources are highly concentrated, the distribution of capability becomes correspondingly unequal. Actors controlling infrastructure do not merely supply technical inputs; they shape the boundaries of innovation, access, pricing, dependency, and market participation. Control over computational resources has become a central locus of economic concentration. \citet{kwonPorter2025} similarly show how exclusive access to machine-learning resources and data can affect innovation and competition. This supports the second proposition:

\paragraph{Proposition 2---Infrastructure--Capability Misalignment} The concentration of AI infrastructure conditions the distribution of technical capability and reinforces asymmetries in technological, economic, and strategic power.

The third misalignment concerns capability and productivity. Increases in AI capability expand the range of tasks that can be automated, supported, or reorganised, but they do not automatically generate aggregate productivity or social value. The translation from capability to productivity depends on absorptive capacity, organisational complementarity, task structure, labour institutions, and governance. \citet{acemoglu2025} emphasises that the macroeconomic effects of AI depend on the share of tasks affected, the scale of real cost savings, and the creation of new tasks. Similarly, \citet{acemogluJohnson2023} argue that technological progress does not automatically generate shared prosperity; it depends on whether institutions favour complementarity or substitution. This supports the third proposition:

\paragraph{Proposition 3---Capability--Productivity Misalignment} Increases in AI capability do not produce proportional productivity gains unless they are accompanied by absorptive capacity, organisational complementarity, and institutional arrangements that direct technological change toward value creation rather than mere task substitution.

The fourth misalignment concerns productivity and societal outcomes. Even where AI generates productivity gains, these gains do not automatically translate into welfare, fair distribution, resilience, or legitimacy. Productivity may be captured by firms controlling platforms, infrastructure, data, or capital. It may coexist with labour displacement, declining job quality, increased dependency, environmental externalities, or loss of institutional trust. Research on work and automation shows that the effects of AI vary depending on whether systems complement or replace human labour and on the kinds of tasks affected \citep{delRioChanona2025}. Augmentation-oriented configurations may preserve professional judgement and support productivity gains \citep{alam2024}, but substitution-oriented trajectories may produce concentrated benefits and social disruption. Recent evidence on AI-exposed occupations also suggests that labour-market effects may be unevenly distributed, with employment declines concentrated among early-career workers in occupations where AI is more likely to automate rather than augment human labour \citep{brynjolfsson2025}. This supports the fourth proposition:

\paragraph{Proposition 4---Productivity--Societal Outcomes Misalignment} Productivity gains generated by AI do not automatically produce welfare, equitable distribution, resilience, or legitimacy; the translation from productivity to societal value depends on governance, labour institutions, distributional mechanisms, and the allocation of externalities.

The fifth misalignment concerns governance itself. As the AI regime accelerates, governance must act across finance, infrastructure, capability, productivity, and societal outcomes. However, regulatory systems often respond more slowly than technological and financial systems evolve. \citet{wang2026} identify persistent regulatory lag in the governance of generative AI, while \citet{emeryXu2025} argue that existing approaches to AI governance may be insufficient for advanced and transnational AI systems. Governance therefore becomes a structural condition for managing misalignments before they become embedded in infrastructure, markets, labour systems, or international dependencies. This supports the fifth proposition:

\paragraph{Proposition 5---Governance--Misalignment Proposition} The greater the speed, autonomy, concentration, and financial intensity of the AI regime, the greater the need for anticipatory governance capable of observing, interpreting, and correcting structural misalignments before they become systemic.

These five propositions and their corresponding governance implications are summarised in \cref{tab:misalignments}.

\begin{table}[!htbp]
\centering
\small
\caption{Structural misalignments in the AI regime and corresponding governance responses.}
\label{tab:misalignments}
\begin{tabularx}{\textwidth}{p{0.21\textwidth}YYY}
\toprule
\textbf{Relationship} & \textbf{Structural misalignment} & \textbf{Expected consequence} & \textbf{Governance response} \\
\midrule
Finance $\rightarrow$ Infrastructure & Infrastructure expansion depends on uncertain future returns & Financial fragility, overinvestment, systemic exposure & Transparency, leverage oversight, realistic productivity assessment \\
Infrastructure $\rightarrow$ Capability & Compute, cloud, data, and supply chains become concentrated & Unequal capability, dependency, rent extraction & Competition policy, compute access, interoperability, infrastructure resilience \\
Capability $\rightarrow$ Productivity & Capability expands faster than organisational adaptation & Automation without proportional productivity & Absorptive capacity, organisational redesign, augmentation-oriented incentives \\
Productivity $\rightarrow$ Societal outcomes & Productivity gains are unevenly distributed & Welfare gaps, dependency, distrust, externalities & Distributional mechanisms, labour transition, social and environmental accounting \\
Governance $\leftrightarrow$ Regime & Institutional adaptation lags behind technical and financial acceleration & Lock-in, systemic risk, legitimacy erosion & Anticipatory governance, observability, controllability, adaptability \\
\bottomrule
\end{tabularx}
\end{table}

These propositions are intended to guide empirical operationalisation by identifying observable relations among investment trajectories, infrastructure concentration, capability diffusion, productivity realisation, and governance latency.

As shown in \cref{tab:misalignments}, the proposed framework shifts the focus of analysis from AI capability alone to the structural relations through which capability is financed, enabled, deployed, and socially evaluated. The relevant object of governance is therefore not the isolated AI system, but the regime-level interaction among finance, infrastructure, capability, productivity, and social outcomes.

\FloatBarrier

\subsection{Structural Power in the AI Regime}

The concept of structural misalignment gains further analytical force when interpreted through Susan Strange's theory of structural power. Strange defines structural power as the power to shape the frameworks within which other actors operate, rather than merely the power to influence particular decisions or outcomes \citep{strange1988, strange1996}. This perspective is especially useful for analysing AI because the most consequential forms of power are not limited to the possession of a specific model, dataset, or regulatory instrument. They concern control over the structures that determine what can be produced, how it is produced, how it is financed, and under what rules.

The first question is what is produced. In the AI regime, production includes not only goods and services, but also computational capacity, models, datasets, platforms, infrastructures, and technical standards. Control over infrastructure conditions which capabilities can be developed and which applications become economically viable. Advanced chips, data centres, cloud platforms, proprietary datasets, and energy-intensive computing systems define the material boundaries of AI production. These infrastructures are embedded in globally distributed and politically contested supply chains involving semiconductors, minerals, data labour, energy systems, deployment infrastructures, and waste streams \citep{muldoon2026}. Their expansion also generates environmental implications, especially through data-centre energy demand and decarbonisation pathways \citep{xiao2025}. From a structural power perspective, infrastructure is therefore not a passive technical substrate. It determines the conditions under which capability can be produced and scaled.

The second question is how AI is produced and deployed. The same technical capability may be integrated into organisations in different ways. It may augment human judgement, support learning, and generate new tasks; or it may be used primarily to automate substitutable work, reduce labour costs, and concentrate decision-making. This direction of technological change is not technologically predetermined. It is shaped by organisational incentives, governance arrangements, labour institutions, and business models. When governance and organisational complementarity are weak, deployment tends to favour short-term efficiency gains and the automation of easily substitutable tasks. When governance supports augmentation, learning, and accountability, capability is more likely to generate socially sustainable productivity. Thus, control over organisational architectures and deployment incentives becomes a form of structural power.

The third question is how AI is financed. Finance shapes the speed, direction, and concentration of the regime. Capital-intensive infrastructure requires large-scale investment, and these investments are increasingly linked to expectations of future productivity, revenue, and strategic advantage. Financial structures can accelerate infrastructure expansion, but they can also externalise risk if expected returns fail to materialise. Actors with privileged access to capital can scale infrastructure more rapidly, influence technological priorities, and capture rents from downstream dependence. In this sense, finance is not merely an enabling condition. It structures the future trajectory of the regime by determining what kinds of AI development are funded, at what scale, and under what return expectations.

The fourth question is under what rules the regime operates. Governance determines who may deploy AI systems, under what conditions, with what obligations, and at what cost. Rules concerning standards, auditing, transparency, liability, competition, procurement, data access, compute governance, and safety requirements do not merely correct harms after deployment. They shape the conditions of production, investment, access, and accountability in advance. From this perspective, governance itself is a form of structural power. It defines the institutional environment within which finance, infrastructure, capability, and productivity interact.

This Strangean interpretation clarifies why structural misalignments are politically significant. A finance--infrastructure misalignment may expose some actors to risk while allowing others to capture returns. An infrastructure--capability misalignment may leave adopting countries dependent on external platforms. A capability--productivity misalignment may generate automation without broad value creation. A productivity--societal outcomes misalignment may concentrate gains while weakening welfare, resilience, or legitimacy. A governance misalignment may leave institutions unable to intervene until dependencies and risks are already locked in.

The AI regime is therefore not only an arena of technological innovation. It is a configuration of structural power in which control over finance, infrastructure, capability, productivity, and rules determines the distribution of opportunities, risks, and dependencies. This is especially relevant for actors outside the technological core. For such actors, the central issue is not simply whether they can adopt AI systems, but whether they can influence the structural conditions under which those systems are financed, built, deployed, governed, and translated into social value. The next section applies this framework to Spain and Chile as two illustrative configurations beyond the technological core.

\section{Illustrative International Configurations: Spain and Chile beyond the Technological Core}\label{sec:spainchile}

The proposed framework can be used to analyse how countries outside the technological core participate in the AI regime without necessarily controlling its generative conditions. This section briefly illustrates the framework through Spain and Chile. The objective is not to provide a full comparative case study, but to show how the F--I--C--P architecture distinguishes AI adoption from structural control. In doing so, the section demonstrates that national AI capacity cannot be assessed only by strategies, applications, or readiness indicators; it must also be assessed by the position countries occupy within global relations of finance, infrastructure, capability, productivity, and governance.

Spain and Chile are useful illustrative configurations because both combine active AI policy agendas with limited control over the frontier layers of the global AI regime. They show that participation in AI cannot be reduced to a binary distinction between technological leaders and passive adopters: countries may adopt AI actively and strengthen governance capacity while remaining dependent on external infrastructures, capital, platforms, chips, cloud systems, and frontier models.

Spain represents a configuration of supranationally embedded adoption. Its AI trajectory is shaped not only by domestic capabilities, but also by its insertion into the European Union's regulatory, infrastructural, and funding ecosystem. Spain benefits from European and national digitalisation instruments, public investment, supercomputing capacity, AI strategy, and EU-level regulatory frameworks, including the broader European agenda on digital sovereignty and trustworthy AI \citep{governmentSpain2024, oecdAI2025}. These elements provide significant governance and institutional advantages. They also support scientific capacity, public-sector innovation, language-oriented AI initiatives, and applied AI development.

However, Spain's position also illustrates the limits of regulatory and institutional embedding. Strong governance capacity does not automatically imply structural control over the global AI regime. Spain remains dependent on external providers for many frontier layers of AI development, including advanced semiconductors, hyperscale cloud infrastructure, proprietary platforms, foundation models, and global AI supply chains. Its central challenge is therefore not simply to increase AI adoption, but to translate European institutional embedding into productive capacity, organisational transformation, public value, resilience, and reduced dependency. This is particularly important because the diffusion of AI across firms, SMEs, public administrations, and sectors remains uneven, and productivity gains depend on absorptive capacity and organisational complementarity rather than on the availability of AI tools alone \citep{europeanCommission2025}.

Chile represents a different configuration: regional leadership under structural dependence. Chile has developed a visible AI policy agenda and occupies a leading position within Latin America in regional assessments of AI ecosystems, governance, talent, research, and adoption \citep{ilia2025}. Its institutional stability, digital connectivity, and policy orientation give it a stronger AI profile than many regional peers. It is therefore not analytically useful to describe Chile simply as a laggard or passive adopter. Rather, Chile illustrates how a country can achieve regional leadership while remaining structurally constrained within the global AI regime.

Chile's constraints are linked to scale, capital, infrastructure, frontier compute, and global value capture. External investment can support connectivity, data centres, cloud services, and digital transformation, but it may also deepen dependency if the highest rents are captured by foreign infrastructure providers, platform owners, or technology firms. Similarly, Chile may develop meaningful applied capabilities, governance expertise, and regional leadership without controlling the full-stack conditions of frontier AI development. Its productivity challenge is therefore to convert AI adoption into local productive development, skills formation, SME diffusion, public value, and institutional resilience. As the OECD has emphasised in relation to Chile's digital transformation, the productivity benefits of digital technologies depend on their diffusion across firms and sectors, and on complementary investments in skills, innovation, and organisational capacity \citep{oecdChile2025}.

The comparison between Spain and Chile highlights three implications of the framework. First, adoption does not equal control. Both countries may expand AI use across firms, public services, and organisations, but adoption alone does not determine who controls infrastructure, models, capital, standards, or value capture. Second, governance capacity does not equal technological sovereignty. Spain benefits from EU regulatory embedding, while Chile has developed a strong regional governance profile; nevertheless, both remain dependent on external technological ecosystems. Third, institutional position changes the form of dependency but does not eliminate it. Spain's dependency is mediated by European integration, funding, regulation, and public scientific infrastructure. Chile's dependency is mediated by regional leadership, external investment, connectivity, and global technology chains.

The main differences between these two illustrative configurations are summarised in \cref{tab:spainchile}.

\begin{table}[!htbp]
\centering
\caption{Spain and Chile as non-frontier configurations in the AI macro-regime.}
\label{tab:spainchile}
\small
\renewcommand{\arraystretch}{1.18}
\begin{tabularx}{\textwidth}{
>{\raggedright\arraybackslash}p{0.18\textwidth}
>{\raggedright\arraybackslash}X
>{\raggedright\arraybackslash}X}
\toprule
\textbf{Dimension} & \textbf{Spain--EU} & \textbf{Chile--Latin America} \\
\midrule

Regime position &
Supranationally embedded adoption: strong regulatory and institutional integration, but limited frontier control. &
Regional leadership under structural dependence: strong regional AI profile, but limited scale and full-stack control. \\

Finance &
Access to European and domestic digital funding; limited capture of frontier AI rents. &
Capacity to attract digital investment; dependence on external capital and potential value leakage. \\

Infrastructure &
Strong scientific and European infrastructure; dependence on global chips, cloud platforms, and foundation-model ecosystems. &
Strong regional connectivity and data-centre relevance; dependence on global supply chains and frontier compute. \\

Capability &
Applied, scientific, public-sector, and language-oriented AI capacity; limited control over frontier models. &
Regional research and governance leadership; limited frontier-scale capability and model control. \\

Productivity &
High potential, but uneven diffusion across firms, SMEs, sectors, and public administrations. &
Productivity opportunity through digitalisation; persistent gaps in skills, SME adoption, and innovation diffusion. \\

Governance &
Strong EU regulatory embedding and national initiatives; limited control over global infrastructure and standards. &
Advanced regional policy profile; constraints in implementation capacity, scale, and influence over global standards. \\

Structural dependency &
Mediated by European integration, global platforms, cloud providers, chips, and frontier-model ecosystems. &
Mediated by external capital, infrastructure providers, cloud platforms, and global technology chains. \\

Governance challenge &
Convert regulatory and infrastructural embedding into productive capacity, resilience, and reduced dependency. &
Convert regional leadership into local productive development, institutional resilience, and functional sovereignty. \\

\bottomrule
\end{tabularx}

\vspace{0.4em}
\begin{minipage}{0.96\textwidth}
\footnotesize
\emph{Note.} The table presents illustrative configurations rather than exhaustive comparative case studies. Its purpose is to show how non-frontier economies may participate actively in the AI regime while remaining structurally dependent on external finance, infrastructure, platforms, models, or standards.
\end{minipage}
\end{table}

As \cref{tab:spainchile} shows, Spain and Chile occupy different institutional positions within the AI regime, but both illustrate the distinction between adoption and structural control. Their comparison confirms that the F--I--C--P framework captures more than national AI readiness or adoption levels. It clarifies how countries are positioned within the structural relations of finance, infrastructure, capability, productivity, and governance. For non-frontier economies, the key question is not only whether AI can be adopted, but whether adoption can be translated into locally meaningful productivity, social value, resilience, and institutional legitimacy under conditions of global dependence.

The purpose of these configurations is not empirical validation, but analytical differentiation: they show how the same regime architecture produces distinct forms of dependency depending on institutional embedding, regional position, and access to finance, infrastructure, and governance capacity.

These illustrative configurations show why the governance of AI must address not only adoption and regulation, but also the structural conditions under which countries access, absorb, and benefit from the AI regime.

The regime-level infrastructure lens further clarifies what is at stake for non-frontier configurations such as Spain and Chile. The relevant question is not only whether these countries adopt AI systems, develop national strategies, or support local innovation, but whether they can access, shape, govern, or build infrastructural configurations capable of linking finance, compute, capability development, organisational absorption, productivity realisation, and societal outcomes. From this perspective, national AI strategies should be evaluated by their capacity to support or govern regime-level AI infrastructures across domains such as science, industry, public administration, health, cybersecurity, and defence, rather than by adoption metrics alone. AI Gigafactories, national compute infrastructures, public AI platforms, sectoral AI ecosystems, and critical-service infrastructures thus become evaluable objects of governance rather than isolated technological projects.

\FloatBarrier

\section{Governing the AI Regime: Alignment, Anticipation, and Evaluation}
\label{sec:implications}

The preceding analysis suggests that the governance challenge posed by AI is not limited to the regulation of individual systems. It concerns the steering of a macro socio-technical regime in which finance, infrastructure, capability, productivity, and societal outcomes interact recursively. Responsible AI principles remain essential for the design and deployment of trustworthy systems, but they are insufficient when the object of governance is a regime rather than a single application. Regime-level governance must also address the structural conditions under which AI is financed, scaled, deployed, and translated into productivity and social value.

\subsection{Governing Structural Misalignments}

Governing structural misalignments requires moving beyond compliance-oriented regulation toward regime-level steering. The task is not only to ensure that AI systems satisfy formal requirements, but to shape the structural conditions under which AI investment, infrastructure expansion, capability development, organisational adoption, and productivity distribution evolve. AI governance therefore becomes a problem of regime alignment: keeping financial expectations, infrastructural capacity, technical capability, organisational absorption, and societal value within institutionally sustainable relations. This view is consistent with recent work on the AI-policy-governance nexus, which shows how AI regulation and AI integration can move governance beyond compliance and toward broader stakeholder accountability \citep{cordeiro2026}.

The first governance priority concerns the finance--infrastructure relationship. AI infrastructure requires large-scale investment in semiconductors, data centres, cloud systems, energy, connectivity, and specialised technical capacity. These investments are often justified by expectations of future productivity, revenue growth, or market dominance. Governance should therefore improve transparency around AI infrastructure investment, leverage, debt exposure, private credit, public support, and concentration of financial risk. Financial supervision should not treat AI infrastructure as a purely technological matter; it should assess whether investment trajectories remain aligned with plausible productivity and revenue scenarios.

The second priority concerns infrastructure--capability concentration. Since advanced AI capability depends on compute, cloud platforms, data, energy, and supply chains, actors controlling infrastructure acquire gatekeeping power over downstream innovation and adoption. Governance should therefore address excessive concentration through competition policy, interoperability requirements, cloud portability, public or shared compute infrastructure, procurement rules, open standards, and supply-chain resilience. The aim is not full technological autonomy, which may be unrealistic for many countries, but functional resilience: the capacity to reduce vulnerability, diversify dependencies, and preserve meaningful access to AI capabilities.

The third priority concerns the capability--productivity relationship. AI capability does not automatically become productivity. Governance should therefore support the organisational and institutional conditions through which capability is translated into durable value. This requires attention to absorptive capacity, skills, data governance, organisational redesign, SME diffusion, worker participation, public-sector capacity, and incentives for augmentation rather than substitution-oriented automation alone. Recent work on data-centric AI governance in European public administration similarly shows that responsible organisational value depends on institutional arrangements for data quality, monitoring, accountability, and implementation rather than on technical capability alone \citep{aguilar2026}. The relevant question is not only whether AI is adopted, but whether its adoption produces sustainable productivity, complementary forms of work, and socially useful new tasks.

The fourth priority concerns the translation of productivity into societal outcomes. Even where AI generates efficiency or output gains, these gains may be unevenly distributed. Productivity may coexist with wage stagnation, labour displacement, dependency on external providers, environmental costs, or declining public trust. Governance should therefore evaluate AI productivity in relation to welfare, distribution, resilience, and legitimacy. This implies the use of distribution-sensitive indicators, job-quality metrics, environmental accounting, transition support, and public-interest obligations. The social value of AI depends not only on what it enables technically, but on how benefits, risks, costs, and responsibilities are institutionally allocated.

\subsection{Anticipatory Governance of Regime-Level AI Infrastructure}

The structural misalignments identified in this article require anticipatory governance. Anticipatory governance is not merely faster regulation; it is the capacity to steer emerging socio-technical trajectories before they become locked into infrastructures, financial obligations, organisational routines, labour-market arrangements, or dependency relations. Within the AI regime, this means identifying interactions among finance, infrastructure, capability, productivity, governance, and societal outcomes before they generate systemic vulnerability.

This concern is especially relevant once concrete AI configurations acquire regime-level infrastructural character. AI Gigafactories, foundation-model platforms, national compute infrastructures, public AI platforms, sectoral AI ecosystems, and corporate AI infrastructures may become difficult to reverse once investment commitments, access rules, organisational routines, technical standards, and dependency relations are stabilised around them. Anticipatory governance therefore concerns not only the regulation of deployed AI systems, but also the early steering of infrastructural configurations before they become locked into financial, organisational, and institutional pathways.

This concern is consistent with path-dependence research, which shows how increasing returns, timing, and institutional sequencing can make early trajectories difficult to reverse once they become embedded in organisational and political arrangements \citep{pierson2000}.

This broader approach depends on four institutional capacities. Observability enables institutions to detect concentration, dependency, financial exposure, labour disruption, productivity gaps, infrastructural bottlenecks, and emerging systemic risks. Interpretability enables them to understand how developments in one dimension affect the others. Controllability provides instruments capable of altering trajectories, including regulation, standards, public investment, competition policy, procurement, auditing, compute governance, financial supervision, infrastructure access rules, and public-interest obligations. Adaptability enables institutions to revise these instruments as technologies, markets, infrastructures, and risks evolve. Governance effectiveness therefore depends not only on the existence of rules, but on the capacity to observe, interpret, influence, and adapt to the evolving regime.

International coordination remains necessary, but it should be understood as a condition for governing regime-level AI infrastructures rather than as a separate governance layer. Even when such infrastructures are nationally or regionally anchored, their enabling conditions are transnational: finance, semiconductors, cloud platforms, data flows, standards, energy systems, supply chains, foundation models, and deployment infrastructures operate across jurisdictions. National governance alone cannot address these dependencies. Coordination is therefore required around shared standards, compute governance, financial transparency, supply-chain resilience, environmental reporting, interoperability, institutional capacity-building, and the participation of non-frontier economies in rule-making processes. Without such coordination, regime-level AI infrastructures may reproduce the asymmetries they are intended to reduce: frontier actors set standards, control critical inputs, capture rents, and export dependency, while adopting economies internalise social, organisational, and regulatory costs.

\subsection{Limitations, Evaluative Utility, and Research Agenda}

The framework proposed in this article is conceptual. Its purpose is to specify the structural relations through which AI becomes a macro socio-technical regime, not to provide a complete empirical validation, a formal dynamic model, or a full evaluative methodology. The discussion of Spain and Chile illustrates analytical applicability, but it does not constitute a comparative case study. Similarly, the discussion of regime-level AI infrastructure clarifies the unit of analysis through which the macro-regime becomes observable, but it does not yet provide a complete assessment protocol for specific infrastructures. Future research should therefore operationalise the framework across countries, sectors, firms, public administrations, and infrastructural systems.

Beyond its conceptual contribution, the framework has potential evaluative utility for the next generation of AI governance. As AI systems become embedded in cloud platforms, organisational workflows, public services, scientific infrastructures, sectoral decision architectures, and national or regional AI strategies, they increasingly function less as isolated applications and more as enabling infrastructures. In such contexts, the relevant question is not only whether an AI system is safe, accurate, or compliant, but whether it creates durable dependencies, concentration, lock-in, governance latency, substitutability problems, and asymmetric capacity to capture value. The F--I--C--P framework can therefore be used as a diagnostic lens for assessing regime-level AI infrastructures before their financial, infrastructural, organisational, and societal effects become difficult to reverse.

Future research could translate this diagnostic lens into evaluative instruments for AI systems, platforms, facilities, and organisational ecosystems that function as regime-level AI infrastructure. Such instruments could include an AI Infrastructure Assessment Framework to determine when AI systems become embedded in critical organisational, sectoral, or public functions; a Structural Misalignment Risk Scorecard to identify tensions among finance, infrastructure, capability, productivity, governance, and societal outcomes; an AI Dependency and Resilience Map to assess reliance on external providers, cloud platforms, compute resources, models, data, standards, and supply chains; a Regime Alignment Dashboard to monitor whether financial expectations, infrastructural capacity, organisational absorption, productivity realisation, and societal outcomes remain within sustainable relations; and an Evaluation Protocol for AI as Infrastructure to guide regulators, public institutions, firms, and international organisations in assessing access, substitutability, interoperability, accountability, resilience, and systemic effects.

Five research directions follow.

First, future work should develop indicators for finance, infrastructure, capability, productivity, governance, and societal outcomes, including indicators of regime-level infrastructural character.

Second, it should analyse dynamic misalignment by examining the different speeds at which financial expectations, infrastructure expansion, technical capability, organisational absorption, productivity realisation, governance capacity, and societal outcomes evolve.

Third, it should apply the framework comparatively across frontier economies, non-frontier developed economies, and emerging or middle-income economies, as well as across sectors such as healthcare, education, manufacturing, public administration, finance, defence, cybersecurity, and science.

Fourth, it should examine which institutional arrangements improve observability, interpretability, controllability, and adaptability, including AI observatories, regulatory sandboxes, public compute, independent auditing, competition authorities, financial supervisors, labour institutions, procurement systems, standards bodies, and international organisations.

Fifth, future research should examine how regime-level AI infrastructures and structural misalignments are transformed under geopolitical rivalry, where infrastructure control, automated scientific capacity, compute chokepoints, supply-chain dependence, and strategic threat perception may convert technological acceleration into a governance problem of international order.

The framework therefore provides a basis for both conceptual and empirical extension. Its immediate contribution is to clarify that AI governance must operate at the level of structural relations and regime-level infrastructures, not only at the level of individual AI systems. Its future development requires measuring those relations, modelling their temporal dynamics, identifying when AI configurations acquire infrastructural character, and determining which forms of governance capacity can act before misalignments become locked in. This requires moving from general principles of responsible AI toward indicators of regime observability, institutional responsiveness, infrastructural resilience, financial exposure, organisational absorption, dependency, substitutability, distributional impact, and societal legitimacy.

\section{Conclusions}
\label{sec:conclusion}

The article's central claim is that AI governance should be understood as regime alignment: the institutional task of keeping financial expectations, infrastructural capacity, technical capability, organisational absorption, productivity realisation, governance capacity, and societal outcomes within sustainable relations. This claim follows from understanding AI not merely as a set of increasingly capable systems, but as an emerging macro socio-technical regime. AI becomes socially and economically consequential when technical capabilities are financed, embedded in infrastructure, deployed through organisations, translated into productive value, and governed within broader institutional and political-economic structures.

The article has developed the F--I--C--P framework to explain this regime architecture. Finance enables infrastructure; infrastructure conditions capability; capability creates the possibility of productivity; and expectations of future productivity feed back into present financial commitments. This recursive structure explains both the transformative potential and the fragility of the AI regime. It shows why technical capability alone cannot determine AI's societal trajectory: capability becomes consequential only through financial, infrastructural, organisational, institutional, and distributive conditions.

A central contribution of the article is to clarify the concrete unit through which the macro-regime becomes observable. The AI regime should not be understood only as an abstract global system. It is instantiated through regime-level AI infrastructures: configurations such as AI Gigafactories, foundation-model platforms, national compute infrastructures, public AI platforms, sectoral AI ecosystems, and corporate AI infrastructures. These configurations become analytically significant when they integrate capital commitments, computational infrastructure, capability production, organisational dependence, productivity expectations, governance arrangements, and societal consequences.

This shift redefines the object of AI governance. The relevant object is not only the individual AI system, nor even AI adoption at organisational or national level, but the structural alignment among capital allocation, material infrastructure, technical capability, productivity realisation, governance capacity, and societal outcomes. Once AI configurations acquire regime-level infrastructural character, they cannot be assessed only through safety, accuracy, or compliance. They must also be evaluated in terms of dependency, concentration, substitutability, lock-in, productivity realisation, governance latency, resilience, and legitimacy.

The central analytical problem is structural misalignment. Its significance lies not only in the existence of gaps among finance, infrastructure, capability, productivity, governance, and societal outcomes, but in the way these gaps become stabilised through investment commitments, infrastructural dependencies, organisational routines, market concentration, and delayed institutional response. Misalignment therefore explains why the governance of AI cannot be reduced to correcting failures in individual systems after deployment. It must also anticipate how regime-level infrastructures may lock in dependencies, asymmetries, and value-capture arrangements before they become difficult to reverse.

The article has therefore reconceptualised governance as a transversal steering capacity acting across finance, infrastructure, capability, productivity, regime-level infrastructures, and societal outcomes. Effective governance depends on institutional capacities for observability, interpretability, controllability, and adaptability. Responsible AI principles remain essential, but they must be complemented by anticipatory governance directed toward structural relations, systemic dependencies, financial exposure, infrastructural concentration, labour transformation, distributional outcomes, and the evaluative conditions under which AI systems become infrastructure.

The illustrative discussion of Spain and Chile showed how the framework can clarify the position of countries beyond the technological core. Both cases demonstrate that adoption, regulatory capacity, and regional leadership do not necessarily imply structural control over AI finance, infrastructure, frontier capabilities, or value capture. For non-frontier economies, the key issue is not only whether AI can be adopted, but whether countries can access, shape, govern, or build regime-level AI infrastructures capable of linking compute, capability development, organisational absorption, productivity realisation, resilience, and societal value under conditions of dependency.

The framework also has future evaluative utility. It can support the development of indicators, diagnostic matrices, dependency maps, regime-alignment dashboards, and assessment protocols for AI systems, platforms, facilities, and organisational ecosystems that function as regime-level AI infrastructure. Such instruments would allow researchers, regulators, public institutions, firms, and international organisations to identify when AI configurations become infrastructural, how they distribute dependency and value, and whether governance capacity is sufficient to intervene before misalignments become locked in.

Future research should operationalise the framework empirically by measuring finance, infrastructure, capability, productivity, governance, regime-level infrastructural character, and societal outcomes across countries, sectors, firms, and public administrations. Particular attention should be given to the different speeds at which these dimensions evolve and to governance latency: the delay between the emergence of structural misalignments and the capacity of institutions to respond effectively. Further work should also examine how regime-level AI infrastructures are transformed under geopolitical rivalry, where infrastructure control, automated scientific capacity, compute chokepoints, supply-chain dependence, and strategic threat perception may convert technological acceleration into a governance problem of international order.

The societal trajectory of AI will depend not only on how capable AI systems become, but on whether institutions can govern the regime-level relationships through which capability is financed, scaled, deployed, embedded, and translated into collective outcomes. The central challenge is therefore not simply to accelerate AI development, nor merely to regulate individual AI systems, but to align the structures that finance, build, deploy, evaluate, and govern AI with socially sustainable forms of productivity, resilience, distribution, and legitimacy.

\section*{Acknowledgements}

This publication is part of the project ``Ethical, Responsible, and General Purpose Artificial Intelligence: Applications In Risk Scenarios'' (IAFER) Exp.:TSI-100927-2023-1 funded through the creation of university-industry research programs (Enia Programs), aimed at the research and development of artificial intelligence, for its dissemination and education within the framework of the Recovery, Transformation and Resilience Plan from the European Union Next Generation EU through the Ministry of Digital Transformation and the Civil Service. This work was also partially supported by Knowledge Generation Projects, funded by the Spanish Ministry of Science, Innovation, and Universities of Spain under the project PID2023-150070NB-I00.

\section*{Declaration of generative AI and AI-assisted technologies in the writing process}

During the preparation of this manuscript, the author used ChatGPT to support linguistic refinement, drafting alternatives, and consistency checking across sections. The author critically reviewed, revised, verified, and approved all content, arguments, references, and final wording, and remains fully responsible for the integrity and accuracy of the manuscript.

\bibliographystyle{plainnat}
\bibliography{Ref-Governance-F-I-C-P}

\end{document}